\documentclass[12pt, amsmath, amssymb, article]{nature}

\usepackage{float}
\usepackage{units}
\usepackage{graphicx}
\usepackage{times}
\usepackage{epstopdf}
\usepackage{color}
\usepackage{amsmath}

\title{Design of compensated ferrimagnetic Heusler alloys for giant tunable exchange bias}

\author{Ajaya K. Nayak$^{1*}$, Michael Nicklas$^1$, Stanislav Chadov$^1$, Panchanana Khuntia$^1$, Chandra Shekhar$^1$, Adel Kalache,$^1$, Michael Baenitz$^1$, Yurii Skourski $^2$, Veerendra K. Guduru$^3$, Alessandro Puri$^3$, Uli Zeitler$^3$, J. M. D. Coey$^4$ \& Claudia Felser$^{1*}$}

\begin{document}

\maketitle

\begin{affiliations}
 \item Max Planck Institute for Chemical Physics of Solids, N\"{o}thnitzer Str. 40, D-01187 Dresden, Germany
  \item Dresden High Magnetic Field Laboratory (HLD), Helmholtz-Zentrum Dresden-Rossendorf, D-01328 Dresden, Germany
  \item High Field Magnet Laboratory, Institute for Molecules and Materials, Radboud University Nijmegen, Toernooiveld 7,
        NL-6525 ED Nijmegen, The Netherlands
\item School of Physics and CRANN, Trinity College, Dublin 2, Ireland
\end{affiliations}

\begin{abstract}

The discovery of materials with improved functionality can be accelerated by rational material design.\cite{Eberhart04} Heusler compounds with tunable magnetic sublattices allow to implement this concept to achieve novel magnetic properties.\cite{Graf11}
Here, we have designed a family of Heusler alloys with a compensated ferrimagnetic state. In the vicinity of the compensation composition in Mn-Pt-Ga, a giant exchange bias (EB) of more than 3~T and a similarly large coercivity are established. The large exchange anisotropy originates from the exchange interaction between the compensated host and ferrimagnetic clusters that arise from intrinsic anti-site disorder.
We demonstrate the applicability of our design concept on a second material, Mn-Fe-Ga, with a magnetic transition above room temperature, exemplifying the universality of the concept and the feasibility of room-temperature applications.
Our study points to a new direction for novel magneto-electronic devices. At the same time it suggests a new route for realizing rare-earth free exchange-biased hard magnets, where the second quadrant magnetization can be stabilized by the exchange bias.

\end{abstract}


Exchange bias corresponds to a shift of the hysteresis loop of a ferromagnet along the magnetic field axis due to interfacial exchange coupling with an adjacent antiferromagnet.\cite{Meiklejohn1,Nogues1} It is employed for a variety of technological applications~\cite{Meiklejohn1,Nogues1,Skumryev1,He1,Wu1,Lage1,Parkin1,OGrady1}, including magnetoresistive read heads and sensors. One of the most important uses of exchange bias is in spin-valve devices, where an artificial antiferromagnet exchange-biased by a normal antiferromagnet allows  magnetoresistance to be developed in a small field, stabilizing the magnetic state against repeated cycling.\cite{Parkin1} Although the origin of  exchange bias is still a subject of discussion, its phenomenology has variously been interpreted in terms of rough ferromagnetic (FM)- antiferromagnetic (AFM) interfaces~\cite{Malozemoff1,Kuch1}, a domain state model~\cite{Milt1} and uncompensated interfacial spins~\cite{Nolting1,Takano1}. Although both FM and AFM subsystems are inseparable parts of an EB system, it is the latter that determines the magnitude of the EB in the system.\cite{Takano1,Kodama1,Ali2} Therefore, it is important to search for an appropriate magnetically compensated material to observe a maximum effect. Besides the EB phenomenon, magnetically compensated materials are proposed to play a vital role in antiferromagnetic spintronics~\cite{Wadley13,Soh11} and all-optical switching devices~\cite{Kimel04}. Single spin superconductivity has even been predicted in a spin-compensated half-metal.\cite{Pickett96} In this letter we present a design scheme for a magnetically-compensated ferrimagnet and demonstrate the relevance of this compensated magnetic state in realizing a giant exchange bias and a large coercivity.


Heusler compounds, which provide the material base for our design scheme, are well known for  their multi-functional
properties, such  as the giant  magnetocaloric effect~\cite{Krenke05}, the field-induced  shape-memory  effect~\cite{kainuma06} and the topological insulating property~\cite{Chadov10}. The story of their success is based on the fact that new materials can be designed in the flexible structure of the Heusler family on the basis of simple rules taking into account the position of the atoms, the number of valence electrons, the degree of atomic disorder  and  the strength of the exchange interactions~\cite{Graf11}. One of the famous examples of valence electron count is the robustness of the Slater-Pauling rule, typically  connected to half-metallicity, in the L2$_1$ cubic Heusler compounds.\cite{Wurmeh06}  Together with the fact that in Heusler compounds Mn develops a strong localized magnetic moment, it can be used as a guide to search for compensated ferrimagnets in the Heusler family with 24 valence electrons. On the other hand, this provides us only a preliminary guide, since the compensated ferrimagnetic state typically cannot be combined with half-metallicity. The classical example is Mn$_3$Ga~\cite{Chadov13}, which is predicted to exhibit a half-metallic state in the L2$_1$ cubic phase, but the cubic phase does not form in the bulk due to a strong electronic instability leading to a tetragonal strain breaking both the compensated and half metallic states. However, either a half-metallic or a compensated state can be recovered by chemical substitution.\cite{Kurt14} Here, we focus on the compensated magnetic state in Heusler materials.

Combining Mn$_3$Ga with another material of opposite net magnetization, Mn$_2$PtGa, forms an excellent starting point to design a zero-magnetization structure based on a compensated ferrimagnetic state. Mn$_2$PtGa consists of two non-equivalent types of Mn, one in the Mn-Ga and another in the Mn-Pt planes of the inverse tetragonal structure (space group $I$-$4m2$)~\cite{Nayak13}. The Mn sitting in Mn-Ga planes possesses a higher magnetic moment due to its more localized nature. It couples antiferromagnetically to the Mn in Mn-Pt planes. This configuration results in a net uncompensated magnetization of about 0.5 $\mu_{\rm B}/{\rm f.u.}$ arising from the larger moment of the Mn in Mn-Ga planes.\cite{Nayak13}
On the other hand, the complete replacement of Pt by Mn to form Mn$_3$Ga(space group $I4/mmm$) results in one Mn in the Mn-Ga and two Mn in the Mn-Mn planes (former Mn-Pt plane). Thus, Mn$_3$Ga displays a net uncompensated magnetization of about 1 $\mu_{\rm B}/{\rm f.u.}$ of opposite sign to that in Mn$_2$PtGa.\cite{Chadov13,Rode13} Combining these two materials then suggests that we can  create a fully compensated magnet with a compensation point for a particular Mn/Pt ratio. This design scheme is schematically depicted in Fig. 1. From first-principles calculations, it follows that the critical composition with the zero magnetization is achieved in the solid solution Mn$_{3-x}$Pt$_{x}$Ga at a Pt content of about ${x_0=0.59}$, which is in good agreement with the experimental findings. On optimizing the Mn/Pt ratio in Mn$_{3-x}$Pt$_{x}$Ga, we always find a small lack of compensation in the material, due to the formation of FM clusters by anti-site disorder. This leads to an exceptionally large bulk EB and a large coercivity. In contrast to an artificial antiferromagnet, which is a thin film structure composed of two ferromagnetic layers separated by a coupling layer~\cite{Parkin1}, here we combine two isostructural ferrimagnetic compounds Mn$_3$Ga and Mn$_2$PtGa to obtain an intrinsically anisotropic compensated magnetic state on an atomic scale in a bulk material.


In order to characterize the magnetic properties of Mn$_{3-x}$Pt$_{x}$Ga we have measured the temperature dependence of the low field magnetization, $M(T)$. We find a systematic increase in the ferrimagnetic N\'{e}el temperature ($T_{\rm N}$) with increasing Mn content as shown in Fig.~2. The irreversibility between the ZFC and FC curves reflects the appearance of coercivity.  We suggest that FM clusters embedded in the compensated host are the source of this irreversibility. NMR measurements confirm that these clusters originate from random swaps between Pt in the Mn-Pt planes and Mn in the Mn-Ga planes (Supplementary Fig. 2). The irreversibility between ZFC and FC $M(T)$ curves increases with increasing magnetic field demonstrating that cooling in higher fields helps the FM clusters to grow in size (Supplementary Fig. 3).



The ZFC magnetization at 5~K shows an unsaturated behavior for magnetic fields up to 14~T (inset of Fig.~2), therefore, higher fields are required in order to gain a better understanding of the magnetic state of the system. Pulsed-field magnetization measurements at 4.2 K up to a field of 60~T are shown in Fig. 3a. For $x=0.5$, 0.6, and 0.7 the hysteresis loops close at a field of $35-40$~T. The non-saturating magnetization up to 60~T reflects the strength of the dominant intersublattice Mn-Mn exchange in the polycrystalline compensated ferrimagnetic host. The hysteresis reveals that there is a FM component of $0.1-0.3~\mu_{\rm B}$ in the compensated host, which may be compared with its ferromagnetic collinear saturation of about $6~\mu_{\rm B}$/f.u.. We further obtain coercive fields of $\mu_{0}H_{\rm C} = 2.2$, 3.6, and 3.0 T at 4.2 K for $x = 0.5$, 0.6 and 0.7, respectively. These values are comparable to the maximum $H_C$ observed in rare-earth-based hard magnetic bulk materials. To investigate the exchange interaction between  the FM  clusters and AFM  host, we have measured FC hysteresis loops. The FC hysteresis loops taken at 4.2~K  in a dc-field of $\pm32$~T are depicted in Fig. 3b. The loops for ${x=0.5}$, 0.6, and 0.7 display a large  shift in the negative field direction indicating the existence of a large unidirectional anisotropy and a large exchange bias. The hysteresis loops close in a field of only ${20-25}$~T for all samples. Interestingly, the FC hysteresis loops  recorded after field cooling  in 15~T  and 25~T follow a similar path. This implies that a cooling field of $\mu_{0}H_{\rm CF}=15$~T is sufficient to saturate the FM moments and a larger cooling  field does not further change the  magnetic state of the sample (inset of Fig. 3). For all samples the FC $M(H)$ loops pass almost through the origin in the third quadrant, resulting nearly same coercivity and exchange bias field. The EB increases monotonically for $\mu_{0}H_{\rm CF}$ up to 10~T and then saturates for further increasing cooling fields. It may be worthwhile to mention that field of 10~T is nearly sufficient to saturate the exchange bias, but the full 32~T is required to fully close the loop.


In Mn$_{3-x}$Pt$_{x}$Ga, we obtain a maximum EB of $H_{\rm EB}= 3.3$~T for $x=0.6$. This value is among the largest EB reported so far. The dependence of the EB and coercivity on the Pt concentration measured after field cooling the sample in 15~T is displayed in the inset of Fig. 4a. Both, the EB field ($H_{\rm EB}$) and the coercive field ($H_C$) exhibit a clear maximum around $x=0.6$. We observe a small increase in $H_{\rm C}$  for the FC loops in comparison to the ZFC loops, as generally found for materials showing an EB effect.\cite{Nogues1,Skumryev1,Leighton1} To demonstrate the temperature dependence of the EB we have measured hysteresis loops at different temperatures after field cooling in 5~T for Mn$_{2.4}$Pt$_{0.6}$Ga and Mn$_{2.5}$Pt$_{0.5}$Ga. The EB shows a monotonic decrease with increasing temperature that vanishes around $T_{\rm N}$, as shown in Fig. 4a.

The extremely large EB at low temperatures in the Mn-Pt-Ga Heusler alloys is a good starting point for achieving large EB in compensated materials.  Therefore, to establish a comprehensive picture we applied our design concept to another Heusler system in the Mn-Ga family, Mn-Fe-Ga, where the ordering temperature of the system can be varied over a wide range to above room temperature by tuning the composition~(Supplementary Fig. 7-Fig. 14). Furthermore, Mn$_2$FeGa is nearly in a compensated magnetic state, whereas, Fe$_2$MnGa is ferromagnetic. By adjusting the composition  we can achieve a material with ferromagnetic clusters in an almost compensated ferrimagnetic host and a N\'{e}el temperature above room temperature. The temperature dependence of the  EB  for Mn$_{1.5}$Fe$_{1.5}$Ga and Mn$_{1.8}$FeGa are shown in Fig. 4b. Mn$_{1.5}$Fe$_{1.5}$Ga exhibits an EB of 1.1~T at 2~K, which is observable up to room temperature. For Mn$_{1.8}$FeGa we find an EB of 1.2~T and 0.04~T at 2~K and 300~K, respectively.

Our experiments show that magnetic compensation of the host is an important requirement to achieve a large exchange bias. We now introduce a model to explain these observations.
Assuming ferromagnetic clusters in the compensated ferrimagnetic host that are coherent with the host and coupled to it via exchange bonds at the interface, we can elaborate the conditions for the exchange bias. We assume that the FM clusters and their surroundings are part of a single crystalline grain with a uniform orientation of the uniaxial magnetocrystalline anisotropy. By applying an external magnetic field which is large  enough  to  rotate the
magnetization in the ferromagnetic clusters,  but small  enough not to rotate the compensated ferrimagnetic host, we find that the EB is well described by

\begin{eqnarray}
\begin{aligned}
  H\approx\frac{12J_{\rm AB}S_{\rm A}S_{\rm B}\left[1-\cos(\theta_{\rm K}-\theta_{\rm B})\right]\cdot L^2n^{2/3}
+K_{\rm B}\sin^2(\theta_{\rm K}-\theta_{\rm B})\cdot L^3n}{\delta S_{\rm A}\cdot\cos\theta_{\rm K}\cdot
    N_{\rm A} + S_{\rm B}\cos\theta_{\rm B}\cdot L^3n}\,.
\end{aligned}
\label{eq:EB-general}
\end{eqnarray}

where $S_{\rm A (B)}$ are the amplitudes of the atomic magnetic moments
in the host (cluster), $\delta{S_{\rm A}}$ accounts for the
imperfect magnetic compensation in the host; $N_{\rm A(B),\,AB}$ are the number of atoms inside the host (cluster) and at the cluster/host interface, respectively; $J_{\rm AB}$ is the exchange coupling constant at the cluster/host interface.  ${K_{\rm A,\,B}>0}$ are the anisotropy constants for the host and the cluster, $\theta_{\rm K,\,A,\,B}$ are the polar angles referring to the  orientation of the anisotropy axis and magnetization of the cluster and of the host, respectively.

In case of small FM clusters (${L^3n\ll N_{\rm A}}$) the key role
is played by the compensation of the host.  The EB field can be expressed as,

\begin{eqnarray}
\begin{aligned}
  H(M)\approx \frac{\alpha}{M+\beta}\,,
\end{aligned}
\label{eq:EB-general-1}
\end{eqnarray}

where $M$ is the total magnetization; $\alpha$ and $\beta$ are  parameters depending on the interface exchange-coupling, the particular form of the interfaces, and the number, distribution and magnetic moments of the embedded ferromagnetic clusters. The theoretical estimate agrees well with the experimental data that show a nearly inverse relationship between the  $H_{\rm EB}$ and magnetization (see Fig.~1a).

To estimate the maximum exchange-bias field Eq.~1 can be expressed in terms of the size of the ferromagnetic cluster ($L$),

\begin{equation}
\begin{aligned}
  H_{\rm max}[{\rm T}] = 0.21\cdot(1/L[{\rm\AA}])\cdot10^{5}+0.54\,,
\end{aligned}
\end{equation}

where the first term appears from the interface exchange coupling and the second, from the magnetic anisotropy (Supplementary). In order to achieve the exchange bias fields of $3$~T, the characteristic FM cluster size must not exceed $L\sim 10^4$~\AA.

A key experimental observation is that at low temperature the exchange bias is equal to the coercivity
in all samples. This can be explained  by considering additional conditions for the magnetization switching. For simplicity, we consider only those grains which are oriented differently from the polar angle, ${\theta_{\rm K}=\pi/2}$.  Such grains exhibit a sharp rectangular-like hysteresis, in which the coercivity corresponds to a regime of fast domain-wall motion. This rapid process is the transition between two states: one, comprising the domain wall (induced by the competition between the strong negative magnetic field and the positive exchange pinning at the host/cluster interface) and the new one, which is magnetically uniform. The elimination of the domain wall occurs for the small external field values (${H<0}$) at which the energies of these two states become comparable. The detailed theoretical interpretation of our findings can be found in
supplementary material. From this model we estimate  that $H_{\rm C}^+$ is only about -1~mT, where the cluster size $L$ is the leading contribution. Taking into account that the measured hysteresis is the statistical sum over all orientations, $H_{\rm C}^+$ will have an even smaller amplitude.


To conclude, we have established a new approach to design compensated ferrimagnetic Heusler alloys. A small lack of compensation allows us to achieve an extremely large EB and matching coercivity. A simple model allows us to quantify the experimental results that show an inverse relationship between the total magnetization and magnitude of the exchange-bias field. The present finding provides new insight into the mechanism of EB and  suggests exploring similar coherent materials that can be implemented in thin film. Furthermore, the achievement of a coercivity of more than 3~T based on exchange anisotropy proposes a new approach to permanent magnet design. Permanent magnets are expected to deliver a large flux density. Their working point, stabilized by coercivity, falls naturally in the second quadrant of the hysteresis loop, because of the demagnetizing field. The present work offers an alternative approach - to stabilize the second quadrant magnetization by exchange bias. However, to realizable in practical applications it must be coupled with a larger magnetization in order to produce a useful flux density. A non-percolating ferromagnetic fraction approaching 30\%, rather than around 3\% in Mn-Pt-Ga and less than 12\% in Mn-Fe-Ga, is needed. The balance between the volume of FM clusters and compensated host needs to be optimized to achieve a large exchange bias as well as a large magnetization. Hence, the present finding is a good starting point for designing rare-earth free hard magnets comprising of two hard magnetic phases, which is different form the conventional exchage spring approach. Our design concept together with the flexibility and huge number of materials crystallizing in the Heusler structure indicate the high potential in designing new materials with a large EB and a large coercivity at room temperature. Finally, the large EB and coercivity in the nearly compensated Heusler alloys may significantly expand the importance of these materials in the physics and material science community due to high potential for antiferromagnetic spintronics.





{\bf Acknowledgements:} We thank J.\,A.\,Mydosh and Erik Kampert for valuable discussions on the present work. This work  was financially supported  by the  Deutsche Forschungsgemeinschaft
DFG  (Projects No.~TP~1.2-A  and No.~2.3-A of  Research Unit  FOR 1464
ASPIMATT) and by the ERC Advanced Grant No. (291472) "Idea Heusler". The
experiments at the High Magnetic Field Laboratory Dresden (HLD) and High
Field Magnet Laboratory Nijmegen  were sponsored by Euro-MagNET II under
the European Union Contract No. 228043.

{\bf Author Contributions} All authors contributed substantially to this work.

{\bf Author Informations} The authors declare no competing financial interests. Correspondence and requests for materials should be addressed to C.F. (felser@cpfs.mpg.de) or A.K.N. (nayak@cpfs.mpg.de).


\begin{figure}
\centering
\includegraphics[angle=0,width=10cm,clip]{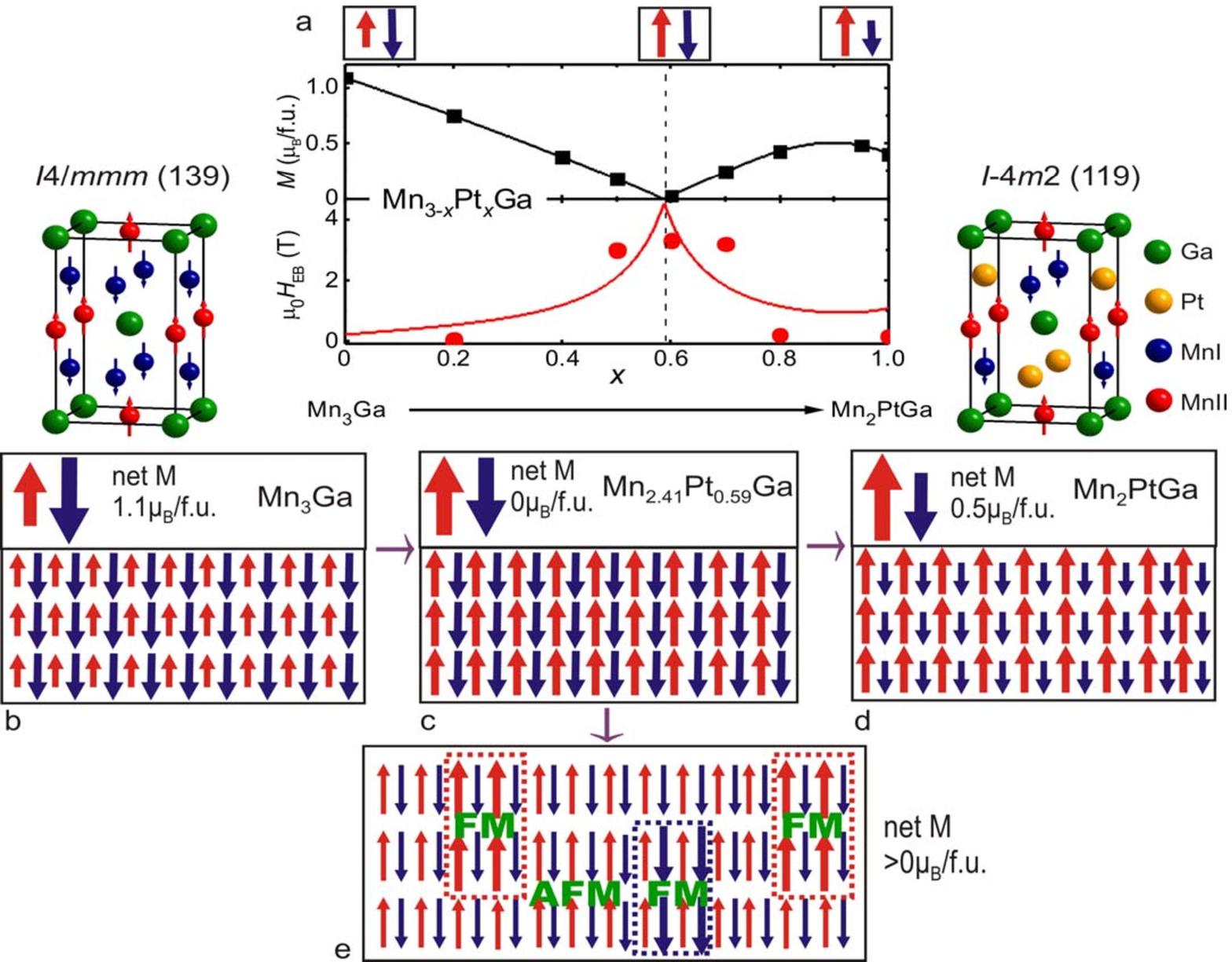}
\caption{{\bf Design of a compensated magnetic state.}
{\bf a}, A theoretical modeling showing that the Pt content controls the amount of Mn-spins pointing down by varying  the uncompensated magnetization, $M$, from ${1.1}~\mu_{\rm B}$ for Mn$_3$Ga to $0.5~\mu_{\rm B}$, in the opposite direction, for Mn$_2$PtGa, thereby, going through the compensation point (${M=0}$) at a Pt content of ${x=0.59}$ (upper panel). Approaching compensation maximizes $H_{\rm EB}$ (lower panel): solid curve corresponds to the model calculation  deduced from the {\it ab-initio} calculated magnetic moments, closed circles  corresponds to the experiment.
{\bf b}, Magnetic moments of manganese in two different sublatices for the tetragonal Mn$_3$Ga (see the crystal structure), one Mn in Mn-Ga plane with a localized moment $\approx 3.1~\mu_{\rm  B}/{\rm f.u.}$ (spin up, red arrow) and two Mn in Mn-Mn planes with a larger sum moment $\approx 4.2~\mu_{\rm  B}/{\rm f.u.}$ (spin down, blue arrow). The Mn in the Mn-Ga plane couples antiferromagnetically to the Mn in the Mn-Mn planes resulting in a net moment of $1.1~\mu_{\rm  B}/{\rm f.u.}$ with spin down in the present illustration.
{\bf c},~Perfect magnetic compensation is expected in Mn$_{2.41}$Pt$_{0.59}$Ga. {\bf d},~Magnetic moment of manganese in two sublatices for Mn$_2$PtGa (see the crystal structure), one Mn in Mn-Ga plane with a larger localized moment (spin up, red arrow) and one Mn in Mn-Pt plane with a smaller moment (spin down, blue arrow). This configuration displays a net moment of around $0.5~\mu_{\rm  B}/{\rm f.u.}$ with spin up. Therefore, on going from  Mn$_3$Ga to Mn$_2$PtGa, a magnetic compensation is encountered with equal and oppositely aligned Mn moments in the Mn-Ga and Mn-Pt planes. {\bf e}, A small anti-site disorder can bring about the formation FM clusters inside the compensated host. The exchange interaction between the FM clusters and the compensated host establishes a strong exchange anisotropy in the system leading to the observation of an exchange bias.}
\label{fig:FIG1}
\end{figure}


\begin{figure}
\centering
\includegraphics[angle=0,width=10cm,clip]{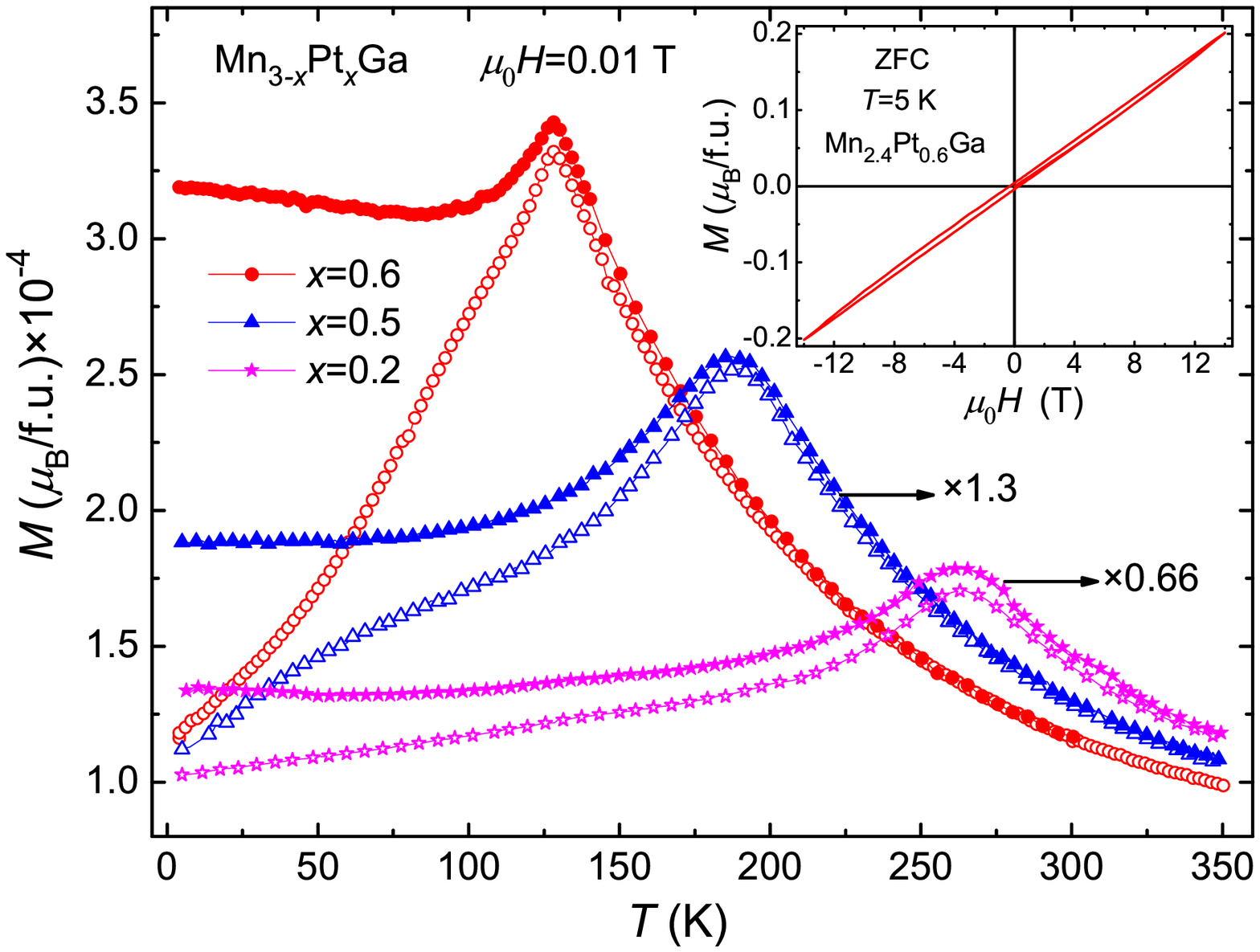}
\caption{{\bf Compensated ferrimagnetic state in  Mn-Pt-Ga.}
$M(T)$ for Mn$_{3-x}$Pt$_{x}$Ga  ($x=$0.2, 0.5  and  0.6) measured  in 0.01~T. Open symbols represent the zero-field cooled (ZFC) curves and closed symbols the field cooled (FC) curves.  In the ZFC mode, the sample  was initially cooled to  2~K and the data  were taken upon increasing the temperature in applied magnetic field. In the FC mode, the data were  collected while  cooling in  field. The inset shows the zero field cooled $M(H)$ hysteresis loop for Mn$_{2.4}$P$_{0.6}$Ga measured at 2~K in a field of $\pm$14~T. }
\label{fig:FIG2}
\end{figure}



\begin{figure}
\centering
\includegraphics[angle=0,width=10cm,clip]{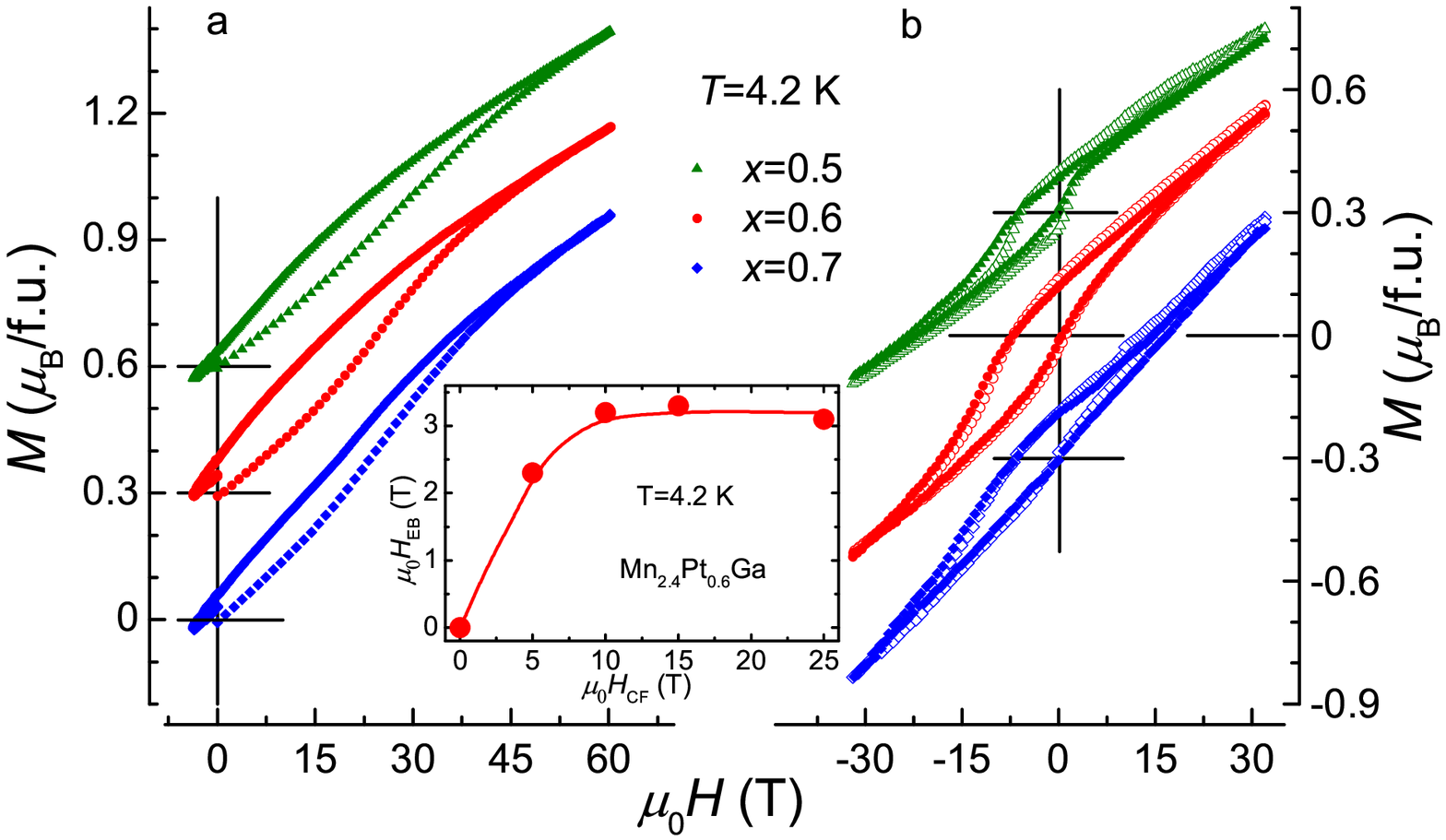}
\caption{{\bf Hysteresis loops in Mn-Pt-Ga.} {\bf a}, $M(H)$ isotherms measured up to 60~T at  4.2~K. The data for successive  samples are shifted  by $0.3~\mu_{\rm B}$ along the magnetization axis for better clarity. {\bf b}, Field cooled (FC) $M(H)$ loops measured up to $\pm~32$~T for $x=0.5$, 0.6, and 0.7 after field cooling the samples in $\mu_{0}H_{\rm CF}=15$~T (closed symbols) and 25~T (open symbols). The loop for  $x=0.7$ is  shifted  by  ${-0.3~\mu_{\rm B}}$  and  for $x=0.5$  by $+0.3~\mu_{\rm B}$  along the magnetization  axis. The inset shows the dependence of  exchange-bias field ($H_{\rm  EB}$) on the cooling field ($H_{\rm  CF}$). $H_{\rm EB}$ is calculated using $H_{\rm EB}=-(H_{\rm C}^{+}+H_{\rm C}^-)/2$, where $H_{\rm C}^+$ and  $H_{\rm C}^-$  are  the lower  and  upper  cut-off  fields at  which  the  magnetization becomes zero.}
\label{fig:FIG3}
 \end{figure}



 \begin{figure}
 \centering
\includegraphics[angle=0,width=10cm,clip]{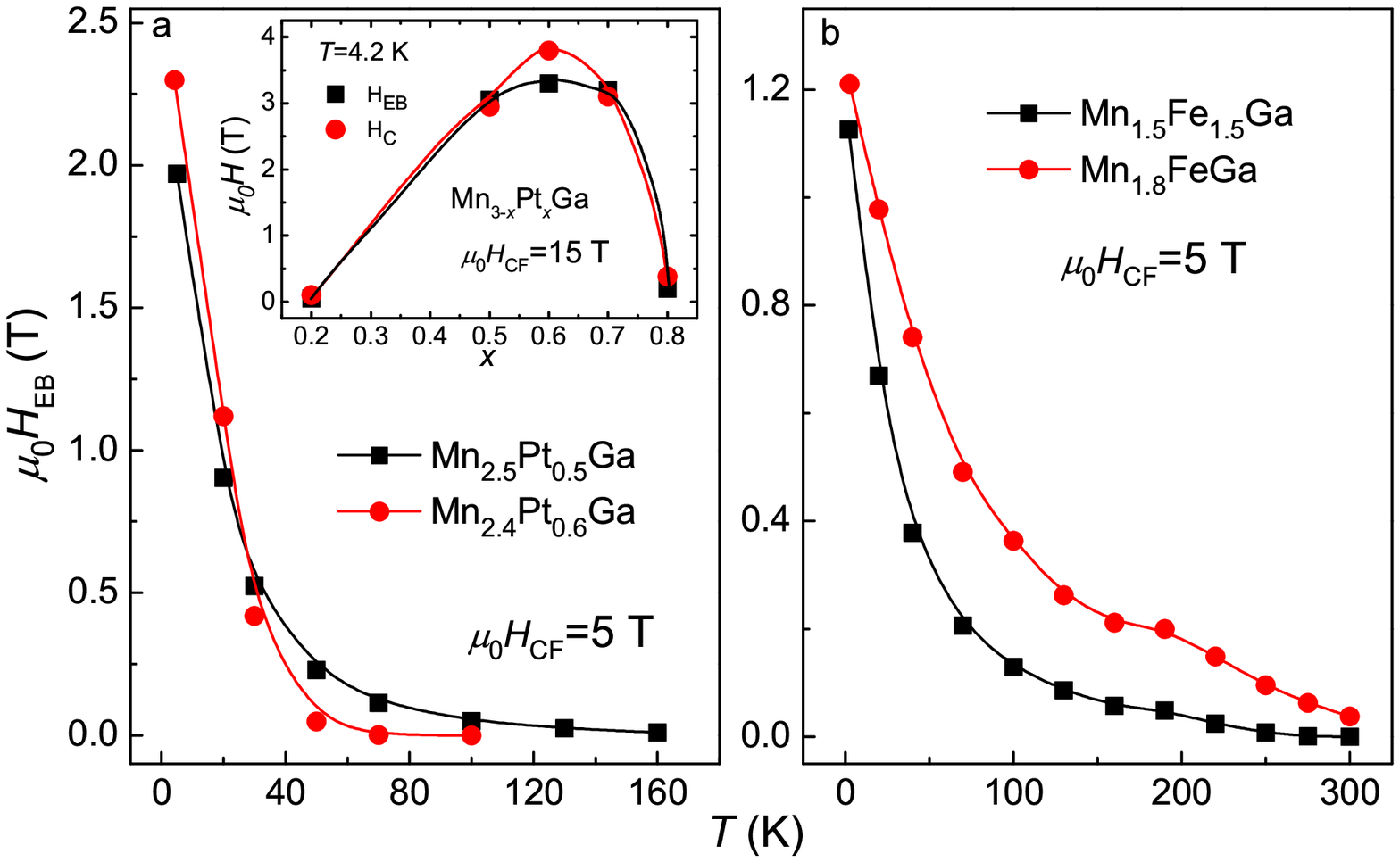}
\caption{{\bf Exchange bias and coercive fields for Mn-Pt-Ga and Mn-Fe-Ga.} {\bf a}, Temperature dependence of the EB for Mn$_{2.4}$Pt$_{0.6}$Ga and Mn$_{2.5}$Pt$_{0.5}$Ga. The  inset of {\bf a} shows  the coercive field  $H_{\rm C}$ (circles) and $H_{\rm  EB}$ (squares) as a function of the Pt concentration $x$ in Mn$_{3-x}$Pt$_{x}$Ga. {\bf b}, Temperature dependence of the EB for Mn$_{1.5}$Fe$_{1.5}$Ga and Mn$_{1.8}$FeGa.}
\label{fig:FIG4}
\end{figure}



\end{document}